\documentclass[prl]{aastex63}
\usepackage{tikz}
\usepackage[english]{babel}
\usepackage[utf8]{inputenc}
\usepackage[colorinlistoftodos, color=green!40, prependcaption]{todonotes}
\usepackage{subfigure}

\begin{document}

\title{Unveiling the nature of galactic TeV sources with IceCube results}

\correspondingauthor{V. Vecchiotti}
\email{vittoria.vecchiotti@ntnu.no}

\author{V. Vecchiotti}
\affiliation{NTNU, Department of Physics, NO-7491 Trondheim, Norway}

\author{F.L. Villante}
\affiliation{University of L'Aquila, Physics and Chemistry Department, 67100 L'Aquila, Italy}
\affiliation{INFN, Laboratori Nazionali del Gran Sasso, 67100 Assergi (AQ),  Italy}

\author{G. Pagliaroli}
\affiliation{INFN, Laboratori Nazionali del Gran Sasso, 67100 Assergi (AQ),  Italy}

\begin{abstract}
IceCube collaboration reported the first high-significance observation of the neutrino emission from the Galactic disk.
The observed signal can be due to diffuse emission produced by cosmic rays interacting with interstellar gas but can also arise from a population of sources. 
In this paper, we evaluate both the diffuse and source contribution by taking advantage of gamma-ray observations and/or theoretical considerations.
By comparing our expectations with IceCube measurement, we constrain the fraction of Galactic TeV gamma-ray sources (resolved and unresolved)  with hadronic nature. 
In order to be compatible with the IceCube results, this fraction should be small or the source proton energy cutoff should be well below the cosmic ray proton knee.
In particular, for a cutoff energy equal to 500~TeV the fraction of hadronic sources should be 
less than $\sim 40\%$ corresponding to a cumulative source flux $\Phi_{\nu, \rm s} \le 2.6 \times 10^{-10} cm^{-2}s^{-1}$ integrated in the 1-100 TeV energy range. This fraction reduces to $\sim 20\%$ for energy cutoff reaching the cosmic-ray proton knee around 5 PeV.
\end{abstract} 



\section{Introduction}


The diffuse galactic neutrino emission produced by hadronic interactions of high-energy Cosmic Rays (CR) with the gas contained in the galactic disk is a guaranteed signal for neutrino telescopes \citep{Pagliaroli:2016,Cataldo:2019qnz,Lipari:2018gzn,Schwefer:2022zly, Evoli2007JCAP}.
The detection of this component is, however, challenging due both to the atmospheric neutrino background and to its subdominant role in all-sky astrophysical neutrino emission \citep{ANTARES:2016mwq,ANTARES:2017nlh,IceCube:2017trr,ANTARES:2018nyb,ANTARES:2022izu}.   
Very recently, IceCube succeeded in its detection thanks to a decade of accumulated statistics and exploiting new machine learning techniques, providing the first detection of the neutrino emission from the galactic plane at the $4.5\sigma$ level of significance \citep{IceCubeScience}. 
%
IceCube exploits a template fitting procedure testing the data compatibility with three models for the expected galactic diffuse neutrino emission.
For each model, the spatial and spectral shapes are frozen to the expected ones while the normalization is free to match the neutrino data considering the entire sky.

All the models considered by IceCube describe the truly diffuse emission expected by CR interactions with the interstellar medium.  
However, freshly accelerated hadrons colliding with the ambient medium within or close to an acceleration site can also produce high-energy neutrinos, see e.g. \cite{Ahlers:2013xia}.
This "sources" component cannot be resolved with the actual statistics and with the poor angular resolution of IceCube cascade events, providing an additional large-scale galactic neutrino emission that adds to the truly diffuse emission due to CR interactions.
The detected IceCube neutrino signal is most likely due to the total galactic neutrino emission where part of the signal could also arise from a population of unresolved point sources, as also stated by the IceCube collaboration. 

In this paper, we discuss the relative importance of truly diffuse and source components by using a multi-messenger approach.
High-energy sources have been observed in the TeV and sub-PeV energy domain by gamma-ray detectors, such as H.E.S.S. \citep{H.E.S.S.:2018zkf}, HAWC \citep{HAWC:2020hrt} and LHAASO \citep{LHAASO:2023rpg}.
%
%
It was recently proven that unresolved gamma-ray sources have a relevant role in the interpretation of the large-scale gamma-ray emission detected in different energy ranges.
In particular, the presence of an unresolved source component at $\sim 10$ GeV summed to the truly diffuse emission can change the spectral shape of the diffuse gamma-ray signal observed by Fermi-LAT mimicking a CRs spectral hardening in the inner Galaxy \citep{Vecchiotti:2021vxp}.   
At very high energy, the presence of the additional diffuse component due to unresolved sources seems needed to obtain a good agreement with the Tibet AS$\gamma$ data, especially at high longitudes \citep{Fang:2021ylv,Vecchiotti:Tibet}.       
All this suggests that sources could give a non-negligible contribution also to neutrino emission in the TeV energy domain explored by IceCube.
The relevance of this component depends, however, on the hadronic or leptonic nature of sources. 
Hadronic processes produce a roughly equal number of charged and neutral pions which decay to neutrinos and gamma rays, respectively. 
This strong correlation between the neutrino and gamma-ray sky, always valid for the truly diffuse emission, fails for the "sources" component if they have a leptonic nature. 
In the following, we discuss the constraints on the fraction of Galactic TeV gamma-ray sources (resolved and unresolved) with hadronic nature that can be obtained from IceCube results.

\section{Results}

The signal observed by IceCube is determined by the total galactic neutrino emission:
\begin{equation} 
\varphi_{\nu,\,{\rm tot}}\left(E_\nu; \, E_{\rm cut}, \xi \right)=\varphi_{\nu,{\rm \,diff}} \left(E_\nu \right) +\varphi_{\nu,\, s} \left(E_\nu; \, E_{\rm cut}, \xi \right)
\label{Ftot_nu}
\end{equation}
which is obtained as the sum of the truly diffuse emission $\varphi_{\nu,\,{\rm diff}}$ produced by CR interactions with the interstellar gas and the cumulative contribution produced by sources $\varphi_{\nu,\,s}$ within a given observation window.
Since sources cannot be individually resolved, the two components cannot be disentangled, unless one uses additional information provided by gamma-ray observations and/or theoretical considerations, as is done in this paper.
The diffuse component can be estimated by using the approach described in~\cite{Pagliaroli:2016,Cataldo:2019qnz},  see App.~\ref{App:diffuse}. 
The obtained predictions depend on the assumed CR spatial and energy distribution, motivating the two cases (labeled as "Case B" and "Case C", respectively) better discussed in the following. 
The cumulative neutrino source flux $\varphi_{\nu,\,s}\left(E_\nu; \, E_{\rm cut}, \xi \right)$ is calculated by using the approach described in \cite{Vecchiotti:2023nqy} which relies on the population study of the sources in the H.E.S.S. Galactic Plane Survey (HGPS) catalog~\citep{H.E.S.S.:2018zkf} performed by \cite{Cataldo:2020qla}, see App.~\ref{App:gamma source} and App.~\ref{App:neutrino source}.
It is obtained by assuming that a fraction $\xi$ of the source population emits gamma-rays and neutrinos due to hadronic interactions of primary nucleon flux $\phi_{\rm p}(E) \propto E^{-\Gamma_{\rm p}} \, \exp\left(-E/E_{\rm cut}\right)$.
In our calculations, the proton spectral index is chosen as $\Gamma_{p} = 2.4$ to reproduce the average spectral properties of HGPS sources while the proton cutoff energy $E_{\rm cut}$ is free to vary.   
The source component is thus obtained in terms of two parameters, $\xi$ and $E_{\rm cut}$, and can be written as:
\begin{equation}
\varphi_{\nu,\, s} \left(E_\nu; \, E_{\rm cut}, \, \xi \right) = \xi \, \Phi^{\rm max}_{\nu, \rm s}(E_{\rm cut}) \,  \phi_\nu (E_\nu;E_{\rm cut})
\end{equation}
where $\Phi^{\rm max}_{\nu,\,s}$ represents the maximal source neutrino flux integrated in the  $\left[ 1, 100 \right]$ TeV energy window, i.e. the neutrino source contribution obtained by assuming that all the TeV gamma-ray sources, resolved and unresolved, are powered by hadronic processes.
 For $E_{\rm cut}\ge 500\,{\rm TeV}$, the maximal source neutrino flux, integrated in the observational window $|b|<5^\circ$ and $0^\circ\le l \le 360^\circ$ considered in this work, is equal to $\Phi^{\rm max}_{\nu,\,s} =  6.4  \times 10^{-10} cm^{-2}s^{-1}$ within 10\% accuracy.
The quantity $\phi_\nu\left(E_\nu; E_{\rm cut}\right)$ is the neutrino spectrum produced by hadronic interactions  (normalized in the same energy window), see App.~\ref{App:neutrino source} for details.
By choosing $\xi=1$ and $E_{\rm cut} = \infty$, we are able to determine the maximal neutrino flux allowed by gamma-ray observation.
It should be remarked that this limit, being based on the entire population of gamma-ray sources, includes by construction also the potential contribution of sources that are not resolved by present gamma-ray telescopes.
A neutrino signal larger than this upper limit can be only obtained by requiring the presence of hadronic source opaque in gamma rays.


%

\begin{figure}[th!]
\begin{center}
\includegraphics[width=0.6\textwidth]{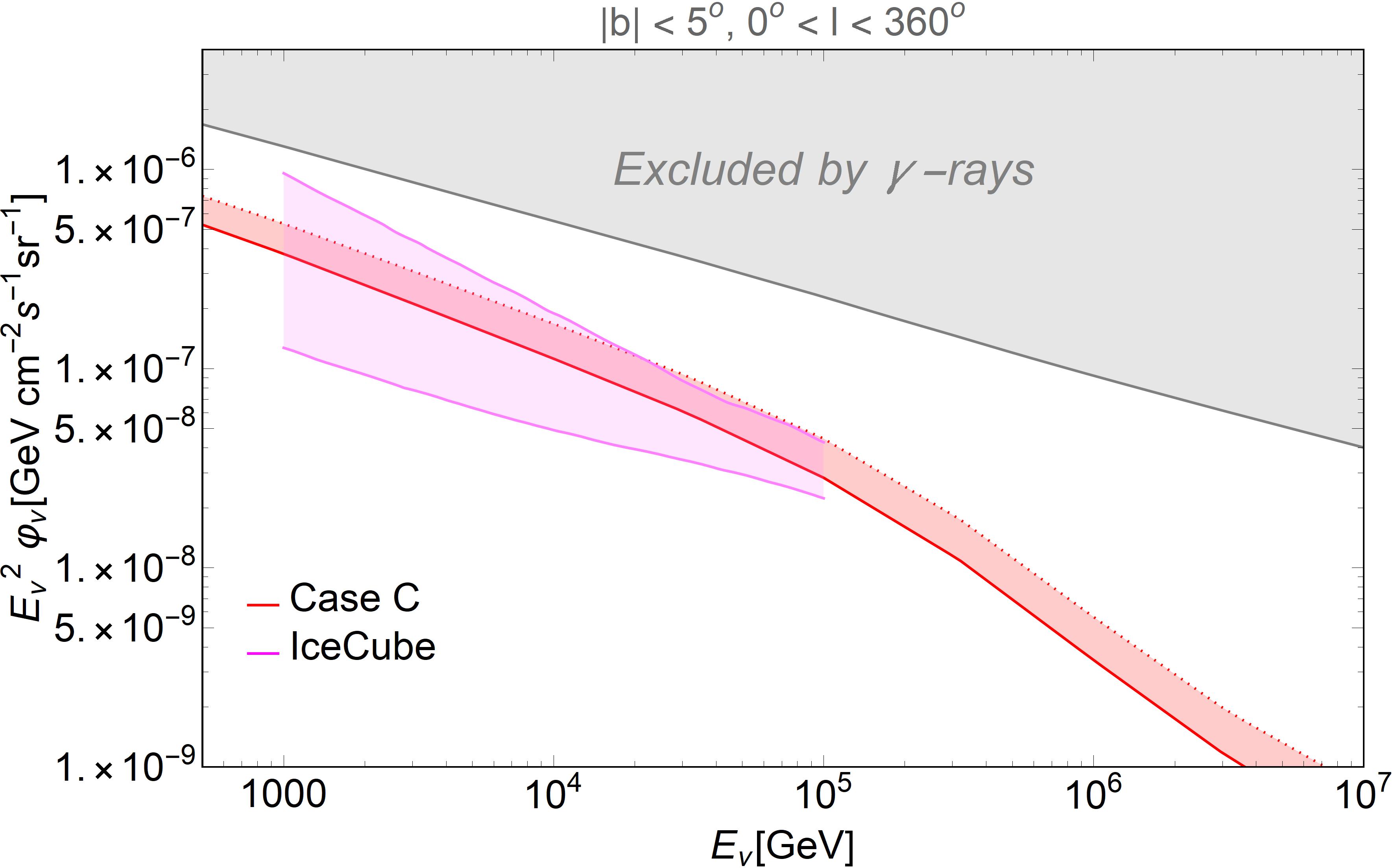}
\caption{\small\em Differential energy spectra of the all flavor diffuse neutrino emission from the Galactic region $|b|<5^\circ$ and $0^{\circ}<l<360^\circ$.  
The magenta region corresponds to the superposition of the three IceCube best-fits for the galactic component with their $1\sigma$ uncertainty. 
The prediction for the diffuse emission (Case C) is shown with a red band. 
The bands represent the uncertainties on the spatial distributions of CRs in our Galaxy. The solid and dotted lines are obtained by assuming smearing radius infinity and 1 kpc, respectively.
We additionally display an excluded region in gray. The bottom line corresponds to the maximum neutrino contribution from our Galaxy obtained by assuming Case C for the diffuse emission, $\xi=1$ and $E_{cut}=\infty$.}
\label{fig:IceCubeC}
\end{center}
\end{figure}
\begin{figure}[h!]
\begin{center}
\includegraphics[width=0.6\textwidth]{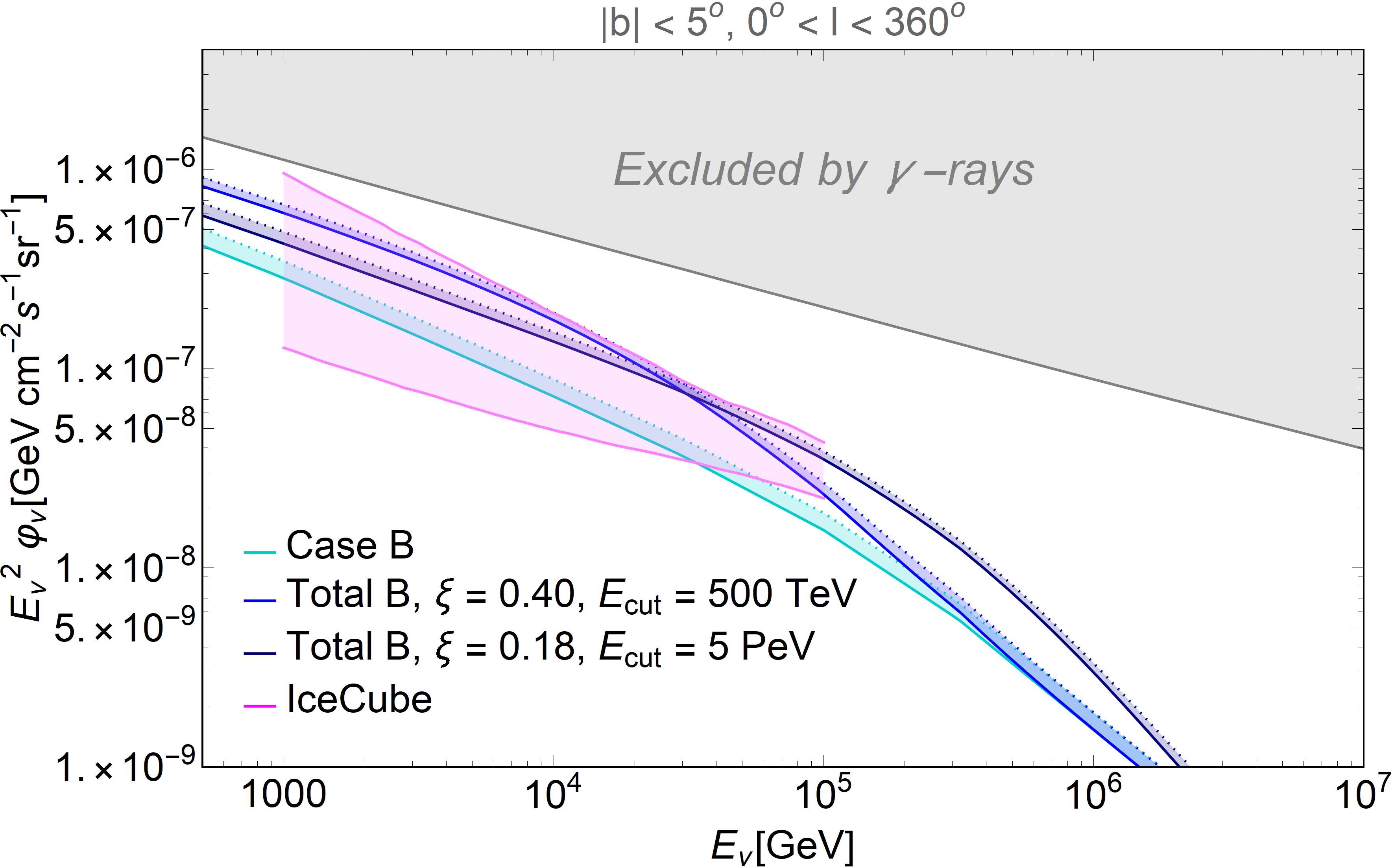}
\caption{\small\em Differential energy spectra of the all flavor diffuse neutrino emission from the Galactic region $|b|<5^\circ$ and $0^{\circ}<l<360^\circ$.  
The magenta region corresponds to the superposition of the three IceCube best-fits for the galactic component with their $1\sigma$ uncertainty. 
The predictions for the diffuse emission (Case B) and the total neutrino flux (Case B + sources) are shown with a cyan band and blue band, respectively. 
The bands represent the uncertainties on the spatial distributions of CRs in our Galaxy. The solid and dotted lines are obtained by assuming smearing radius infinity and 1 kpc, respectively.
We show the effect of different energy cutoffs for the CR source spectra as displayed in the labels.
We additionally display an excluded region in gray. The bottom line corresponds to the maximum neutrino contribution from our Galaxy obtained by assuming Case B for the diffuse emission, $\xi=1$ and $E_{cut}=\infty$.}
\label{fig:IceCubeB}
\end{center}
\end{figure}

In Fig.~\ref{fig:IceCubeC} and Fig.~\ref{fig:IceCubeB} we compare our predictions for the galactic neutrino emission with the IceCube results.
The IceCube galactic signal is obtained by using a template fitting procedure where the angular and energy dependence of the neutrino flux is fixed according to three different models, namely the  $\pi_0$ \citep{Ackermann_2012}, KRA$_\gamma^5$ and KRA$_\gamma^{50}$ models \citep{Gaggero:2015xza}, while the overall normalization is free to vary.
%
%
%
We restrict our comparison to the angular region $0^\circ\le l \le 360^\circ$ and $|b|< 5^\circ$ where the best-fits of the Galactic neutrino component obtained for the different templates give almost the same constraints above $\sim 50$ TeV.
Moreover, in order to be conservative and to take into account the systematic uncertainty related to the adopted template, we show with the magenta region 
the superposition of the regions obtained by IceCube by using different assumptions (including also $1\sigma$ uncertainties of the respective fits).
The displayed band shows that the energy region most effectively probed by IceCube is $50\le E_\nu \le 100\,{\rm TeV}$ since different assumptions basically lead to the same reconstructed flux.
At lower energy, the extracted signal depends instead on the assumed neutrino spectrum. 
In this respect, we recall that the neutrino spectral index is assumed to be equal to 2.7 in the $\pi_0$ model while it is close to 2.5 for the KRA$_\gamma$ models. 
We finally note that the IceCube signal is always below the maximal limit allowed by $\gamma-$ray observations discussed in the previous paragraph (gray solid lines in Fig.~\ref{fig:IceCubeB} and Fig.~\ref{fig:IceCubeC}). 
%
This is a relevant conclusion, different from what obtained by ANTARES \citep{ANTARES:2022izu} that reported a hint for a Galactic neutrino signal which can extend well above this limit, see \cite{Vecchiotti:2023nqy}. 



The truly diffuse neutrino emission $\varphi_{\nu,\,{\rm diff}}$ due to CR interactions with the ISM is displayed by the cyan band in Fig.~\ref{fig:IceCubeB}, labeled as Case B, and by the red band in Fig.~\ref{fig:IceCubeC}, labeled as Case C. 
We calculate this contribution by following the prescriptions of \cite{Cataldo:2019qnz, Pagliaroli:2016} (the details are reported in App.~\ref{App:diffuse}). 
%
The main source of uncertainty for the calculation of this component is the determination of the differential CR flux $\varphi_{\rm CR}\left(E,\,{\bf r}\right)$ as a function of the energy and position in the Galaxy. 
In our Case B, CRs are assumed to have the same spectrum in the entire Galaxy; the flux $\varphi_{\rm CR}\left(E,\,{\bf r}\right)$ can be thus directly linked to its local determination $\varphi_{\rm CR,\,\odot}\left(E\right)$ parameterized by \cite{Dembinski:2017} by a position-dependent normalization factor that is calculated by assuming isotropic diffusion from the (non-uniform) distributions of CR sources in the Galaxy.
The obtained results depend on the adopted diffusion radius $R$.
The upper limit (both for Case B and Case C subsequently discussed), is obtained by taking 
$R = 1$~kpc, i.e. by assuming that CRs are confined relatively close to their sources.
The lower limit is obtained by assuming $R = \infty$ that corresponds to a CR spatial distribution very close
to that predicted by the GALPROP code.
Finally, Case C implements as an additional ingredient the possibility, recently emerged from the analysis of Fermi-LAT gamma-ray data at GeV energies \citep{Pothast:2018bvh, Yang:2016jda, Acero:2016qlg}, that CRs have a harder spectrum in the inner Galaxy than at the Sun position, see App.~\ref{App:diffuse} for details.
As a result of these assumptions, one expects a larger neutrino emission in the TeV domain, with a harder spectral index (that also depends on the direction of the observation), as it is displayed by the red band in Fig.~\ref{fig:IceCubeC}.
%

%
%
%
%



\begin{figure}[h!]
\begin{center}
\includegraphics[width=0.6\textwidth]{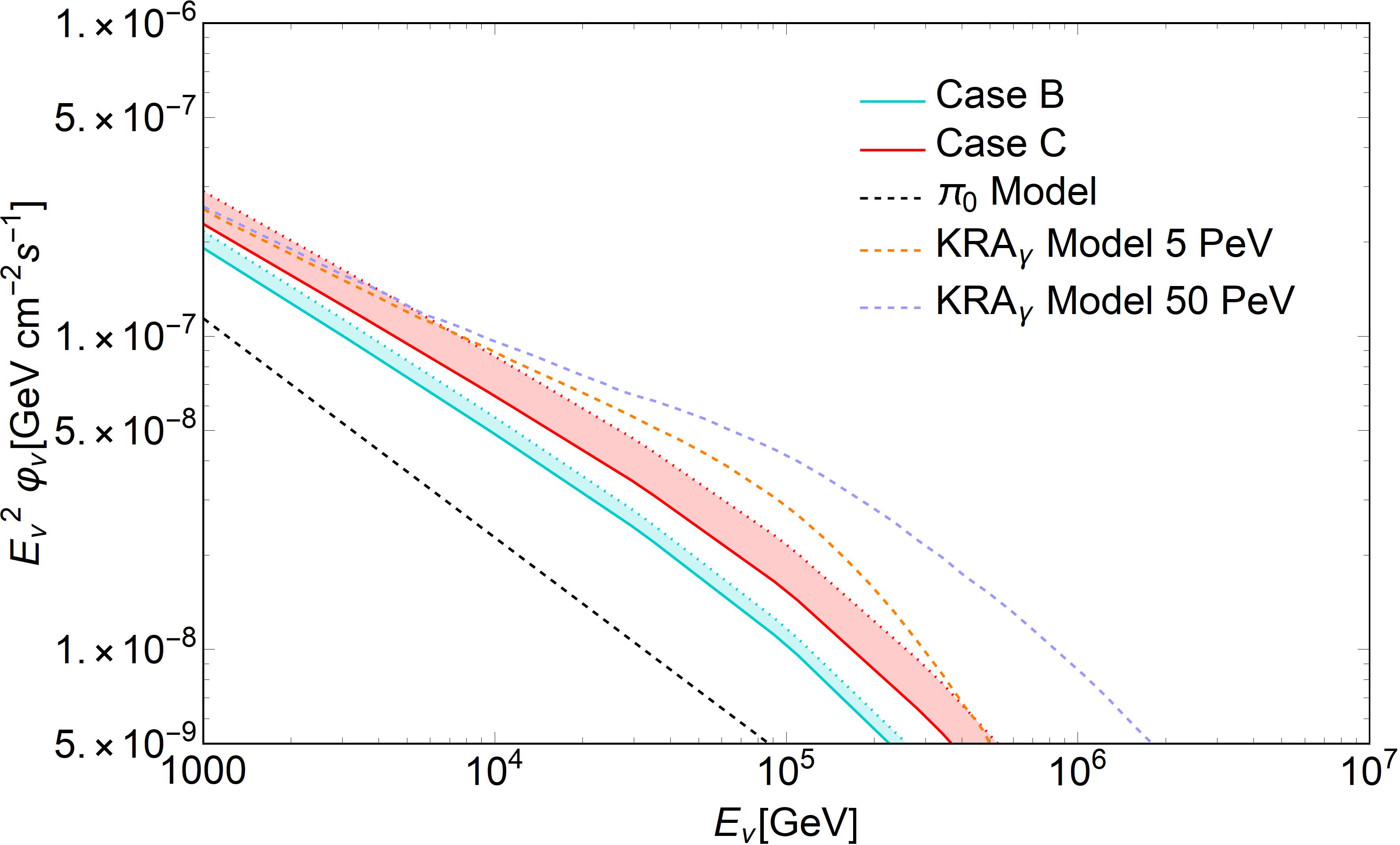}
\caption{\small\em We compare the all sky, single flavor expectations provided by different models for the truly diffuse emission due to the CR interactions with the interstellar medium. The diffuse emission due to Case B and Case C are plotted with a blue and a red band, respectively. Predicted nominal values for the $\pi_0$, the KRA$^5_\gamma$ and the KRA$^{50}_\gamma$ are also displayed for a comparison.}
\label{fig:Diffuse}
\end{center}
\end{figure}

Our predictions for the truly diffuse emission are compared with the three reference models used by IceCube in Fig.~\ref{fig:Diffuse}. 
As it is expected, our Case C is very similar to the KRA$^5_\gamma$ while Case B predicts a diffuse emission which is a factor $\sim 2$ greater than the $\pi_0$ model.
This is due to the fact that the $\pi_0$ model is obtained by extrapolating the neutrino diffuse emission at GeV energies (estimated from gamma-ray data) with a spectral index equal to 2.7.
This is, however, not consistent with the observed CR spectral behavior that
shows a hardening at rigidity $\sim 300$ GV \citep{Adriani_2011,2015AMSp,2015AMSHe}.
This feature is automatically implemented in our calculations but is not considered in the $\pi_0$ model that consequently underestimates neutrino diffuse emission.

The first conclusion that is obtained from our calculations is that the Galactic gamma-ray source population cannot be entirely powered by hadronic mechanisms. 
Indeed, the total predicted neutrino flux that is obtained by taking $\xi=1$ greatly exceeds the IceCube signal both in Case B and Case C, unless the proton cutoff energy is much lower than 100 TeV, i.e. a value that is not compatible with the fact that gamma-ray sources have been observed to emit up to sub-PeV energy domain, see e.g. \citep{HAWC:2019tcx,LHAASO:2023rpg}.
%


This conclusion is particularly strong and rich in physical implications when we consider our Case C, i.e. if we assume that the CR spectral index is position-dependent and becomes harder toward the Galactic center, as obtained from the analysis of the Fermi-LAT data by~\cite{Acero:2016qlg, Yang:2016jda, Pothast:2018bvh}.
Indeed, as it is reported in Fig.~\ref{fig:IceCubeC}, the diffuse emission in our Case C saturates the IceCube signal, leaving no space for any other additional contribution. 
This result is consistent with the best-fit normalization smaller than 1 that was obtained by IceCube analysis for the KRA$^5_\gamma$ model \citep{IceCubeScience}. 
The above result automatically implies that the source contribution to the observed signal should be zero or negligible. 
In other words, one is forced to require that either $\xi\ll 1$ or $E_{\rm cut} \ll 500$~TeV in such a way that the source contribution in the energy range probed by IceCube becomes much smaller than the CR diffuse emission. 
This request, however, could be not easily fulfilled in the context of the model that we are considering.
%
Indeed, CRs up to the proton knee energy are believed to have a Galactic origin.
This implies the existence of sources in our Galaxy that should accelerate hadrons up to few PeVs energy. 
As an example, the KRA$^5_\gamma$ model assumes that the source injection spectrum is a power law with an exponential cutoff at 5~PeV.
In order to not exceed the IceCube signal, one is 
forced to assume that these sources 
accelerate hadrons up to few PeVs but do not effectively produce neutrinos in the 1-100 TeV energy range. 
%

The situation is quite different if we consider our Case B, i.e. we assume that the CR spectrum is uniform within the Galaxy and corresponds to that measured at the Earth and parameterized by \citep{Dembinski:2017}.
In this case, the IceCube data allow for a non-vanishing source contribution that seems to be even required if we restrict the comparison to the most constrained energy range $50\le E_\nu\le 100$~TeV.

The blue bands reported in Fig.~\ref{fig:IceCubeB} show the total (diffuse + sources) neutrino emission evaluated by using  Eq.~\ref{Ftot_nu} and considering selected values of the two parameters $\xi$ and $E_{\rm cut}$.
We see that the allowed fraction $\xi$ of TeV gamma-ray sources that can have hadronic nature depends on the assumed proton cutoff energy. 
If we require that Galactic sources accelerate protons up to the "knee" around $E_{\rm cut} = 5\,{\rm PeV}$ \citep{LIPARI2020102441}, which could represent the end of the galactic CRs component \citep{Gabici:2019jvz}, the maximal fraction is $\sim 20\%$, corresponding to a source contribution integrated between 1 and 100 TeV that is equal to $\Phi_{\nu,\, s} = 1.1\times 10^{-10} cm^{-2}s^{-1}$.
For a smaller cutoff energy $E_{\rm cut} = 500\,{\rm TeV}$, we obtain $\xi \le 40\%$, corresponding to $\Phi_{\nu,\, s} \le 2.6\times 10^{-10} cm^{-2}s^{-1}$.
Larger values for $\xi$ require smaller proton cutoff energies that, however, would correspond to the assumption that the neutrino (and gamma-ray) source emission spectrum is suppressed above ${\rm few}\times 10 \,{\rm TeV}$, with potential difficulties to explain the IceCube signal in the most constrained energy region above 50 TeV.
Finally, we can compare our findings with our present knowledge of TeV gamma-ray sources. 
If we consider the HGPS catalog, we obtain that the cumulative gamma-ray flux integrated in the 1-100 TeV energy range that is produced by potential hadronic sources, i.e. 8 Supernova Remnants and 8 Composite Sources, is about $\sim 12\%$ of the total gamma-ray signal $\Phi_{\gamma,\, s} = 4.2 \times 10^{-10} cm^{-2}s^{-1}$  produced by the entire (resolved + unresolved) source population (see tab.~1 of \cite{Cataldo:2020qla}).
Converted in neutrinos, these 16 sources would account for a cumulative flux at a level of $\sim 6.0  \times 10^{-11} cm^{-2}s^{-1}$.
This flux is not negligible and compatible with our limits for Case B, thus potentially confirming this scenario in which a comparable contribution to the IceCube signal is provided by diffuse and source components and disfavoring instead our Case C which requires a negligible source contribution.
However, the number of identified sources of this kind is still very limited not allowing us to reach this conclusion on firm statistical grounds.

\section{Summary}



In conclusion, we have discussed the implications of the recent measurement of high-energy neutrino emission from the Galactic disk performed by IceCube. 
We have shown that the IceCube signal is compatible with the upper limit allowed by TeV gamma-ray observations calculated by \cite{Vecchiotti:2023nqy}. 
Moreover, we have demonstrated that only a fraction of the TeV-Galactic gamma-ray sources can have hadronic nature. 
This fraction has to be negligible if we assume that CRs diffusing in the inner Galaxy have a spectrum harder than at the Sun position, as it is e.g. assumed in the KRA$\gamma$ models or, equivalently, in our Case C. 
This may not be compatible with the fact that these models require the existence of sources
in our Galaxy that accelerate hadrons up to few
PeVs.
Moreover, the observed gamma-ray sources with potential hadronic nature in the HGPS catalog (i.e. SNRs and Composite sources) already account for a non-negligible flux $\sim 10\% \, \Phi_{\nu, \, \rm s}^{\rm max}$ when converted in neutrinos.

If we consider instead the standard scenario in which the CR spectrum is uniform within the Galaxy (i.e. Case B), the maximally allowed fraction is $\xi\le 40\%$, for a cutoff energy of the source proton spectrum $E_{\rm cut}=500$~TeV, corresponding to a cumulative source flux from the Galactic plane $\Phi_{\nu, \, \rm s} \le 2.6 \times 10^{-10} cm^{-2}s^{-1}$.
%
Lower cutoff energies are not consistent with the IceCube signal at $\sim 100$ TeV while
larger cutoffs lead to smaller values for $\Phi_{\nu, \, \rm s}$.
In particular, the fraction of hadronic Galactic sources compatible with IceCube results reduces to $\sim 20\%$ for energy cutoff reaching the cosmic-ray proton knee around 5 PeV.

\section{Acknowledgements \label{sec:acknowledgement}}
The work of VV is supported by the European Research Council (ERC) under the ERC-2020-COG ERC Consolidator Grant (Grant agreement No.101002352).
The work of GP and FLV is partially supported by the research grant number 2017W4HA7S ''NAT-NET:
Neutrino and Astroparticle Theory Network'' under the program PRIN 2017 funded by the Italian Ministero dell'Istruzione, dell'Universita' e della Ricerca (MIUR).

\vspace{0.5cm}

\appendix

\section{Neutrino diffuse emission}
\label{App:diffuse}

%
The neutrino diffuse flux is calculated following the approach of \cite{Pagliaroli:2016, Cataldo:2019qnz} that is summarized in the following.
The differential one-flavor neutrino flux can be parametrized as:
\begin{equation} 
\varphi_{\nu,{\rm diff}}(E_{\nu},\hat{n}_{\nu}) =
\frac{1}{3} \sum_{l=e,\mu,\tau}  \int_{E_{\nu}}^{\infty} dE\; \frac{d \sigma_{l}(E,E_{\nu})}{dE_{\nu}} \int_{0}^{\infty} dl\; \varphi_{CR}(E ,r_{\odot} + l \hat{n}_{\nu})\, n_{\rm H}(r_{\odot} + l\hat{n}_{\nu}) 
\label{diffuse flux}
\end{equation}
where $E_{\nu}$ and $\hat{n}_{\nu}$ indicate respectively the neutrino energy and arrival direction, while $\frac{d \sigma_{l}(E, E_{\nu})}{dE_{\nu}}$ represents the differential cross section for the production of neutrino and antineutrino with flavor $l$ by a nucleon of energy $E$ in a nucleon-nucleon collision. 
In Eq.~\ref{diffuse flux}, the neutrino flux at Earth is assumed to be equally distributed among the different flavors due to neutrino mixing (see, e.g., \cite{Palladino:2015}). 
The nucleon-nucleon cross-section is parameterized by using \cite{Kelner:2006tc}.
The number density of target nucleons $n_{\rm H}({\bf r})$ contained in the gas is taken from the GALPROP code\footnote{GALPROP is made available at \url{https://galprop.stanford.edu/}}, and includes the contributions from atomic $\rm{H}$ and molecular $\rm{H}_{2}$ hydrogen \citep{2002ApJ...565..280M}.
We take into account the contribution of heavy elements by assuming that the total mass of the ISM is a factor $1.42$ larger than the mass of hydrogen \cite{Ferriere:2001rg}.
The differential CR flux $\varphi_{CR}(E ,{\bf r})$ can be written as:
\begin{equation}
\varphi_{\rm CR}(E,{\bf r}) = \varphi_{\rm CR,\odot}(E)\,g({\bf r})\,h({E,\bf r})
\label{Eq:CR_flux}
\end{equation}
where $\varphi_{\rm CR,\odot}(E)$ represents the local nucleon flux which is described according to the data-driven parameterization provided in \cite{Dembinski:2017}.

The function $g(\bf r)$ describes the spatial distribution of CRs and is an adimensional function (normalized to one at the Sun position ${\bf r}_\odot=8.5$ kpc). It is obtained as the solution of a 3D isotropic diffusion equation with constant diffusion coefficient and stationary CR injection $f_{\rm S}({\bf r})$: 
\begin{equation}
     g({\bf r}) = \frac{1}{\rm N}\;\int d^3 x\; 
     f_{\rm S}({\bf r}-{\bf x})\;
     \frac{{\mathcal F}(|{\bf x}|/R)}{2\pi |{\bf x}|}
     \label{Eq:g_funct1}
\end{equation} 
where $f_{\rm S}({\bf r})$ is assumed to follow the SNR number density parameterization given by \cite{Green:2015isa} and $\rm{N}$ is a normalization constant:
\begin{equation}
     \rm{N} = \int d^3 x\; 
     f_{\rm S}({\bf r}_{\odot}-{\bf x})\;
     \frac{{\mathcal F}(|{\bf x}|/R)}{2\pi|{\bf x}|}
\end{equation} 
while the function ${\mathcal F}(\delta)$ is defined as:
\begin{equation}
  {\mathcal F}(\delta) \equiv \int_{\delta}^{\infty} d\gamma\; \frac{1}{\sqrt{2\pi}}\,\exp{\left(-{\gamma}^2/2\right)}
\end{equation} 
The solution depends on the diffusion length $R$, for which we assume two extreme values, $R=1$ kpc, and $R=\infty$ that allow us to reproduce the behavior of the CR density at $E\sim 20\,{\rm GeV}$ obtained by analysis of Fermi-LAT data, see \cite{Cataldo:2019qnz} for details. 

The function $h({E,\bf r})$  introduces the possibility of a position-dependent CR spectral index as inferred from analysis of the Fermi-LAT data (see, e.g., \citep{Acero:2016qlg, Yang:2016jda, Pothast:2018bvh} and it is defined as:
\begin{equation}
h(E,{\bf r})=\left(\frac{E}{\overline{E}}\right)^{\Delta({\bf r})}
\label{Eq:h_funct}
\end{equation}
where $\overline{E}=20\,{\rm GeV}$  is the pivot energy and $\Delta({\bf r}_\odot)=0$. 
The function $\Delta({\bf r})$ in Galactic cylindrical coordinates is modeled as:
\begin{equation}
\Delta(r,z)=\Delta_0\left(1 - \frac{r}{r_{\odot}} \right)
\end{equation}
for $r\le 10$~kpc, while it is assumed to be constant for larger distances. The factor $\Delta_0=0.3$ represents the difference between the CR spectral index at the Galactic center and its value at the Sun position.

\section{Total gamma-ray source flux}
\label{App:gamma source}

The cumulative $\gamma$-ray source signal is calculated following the approach of \cite{Cataldo:2020qla}.
The source spatial and luminosity distribution is described as:
\begin{equation}
\frac{dN}{d^3 r\,dL} = \rho\left({\bf r} \right) Y \left(L\right)  
\label{SpaceLumDist}
\end{equation}
where ${\bf r}$ indicates the source position and $L$ is the source $\gamma$-ray intrinsic luminosity in the $1-100\, {\rm TeV}$ energy range probed by the H.E.S.S. detector. 
The spatial distribution $\rho({\bf r})$, normalized to one when integrated over the entire Galaxy, is proportional to the pulsar distribution parameterized by \cite{Lorimer:2006qs} and scales as $\exp \left(-\left|z  \right|/H\right)$ with $H=0.2\ {\rm kpc}$, along the direction $z$ perpendicular to the Galactic plane.
The source luminosity function $Y(L)$ is described by:
\begin{equation}
Y(L)=\frac{\mathcal N}{L_{\rm max}}\left(\frac{L}{L_{\rm max}}\right)^{-\alpha}
\label{LumDist1} 
\end{equation}
in the luminosity range $L_{\rm min}\le L \le
L_{\rm max}$.
In the above relation, $L_{\rm max}$ and $\mathcal N$ are the maximum TeV $\gamma$-ray luminosity of the population and the high-luminosity normalization of the luminosity function, respectively.
The total TeV $\gamma$-ray flux produced by all the sources (resolved and unresolved) in a given observational window (OW) is calculated by using the prescription of \cite{Cataldo:2020qla}:
\begin{equation}
\Phi_{\gamma,\, \rm S} = 
\frac{ \mathcal N F_{\rm max}}{4\pi (2-\alpha)}\; 
\int_{\rm OW}d^3r \, \rho({\bf r}) \; r^{-2}
\label{phitot}
\end{equation}
where $F_{\rm max} = L_{\rm max}/\langle E \rangle$ represents the maximum TeV emissivity, and $\langle E \rangle=3.25$ TeV is the average energy of photons emitted in the range $1-100\,{\rm TeV}$ obtained by assuming that all the $\gamma$-ray sources have a power-law spectrum with a spectral index equal to $2.3$ \citep{H.E.S.S.:2018zkf}.
The best-fit value of $\mathcal{N}$ is not sensible to a change in the spectral assumption while $L_{\rm max}$ is shifted proportionally to the variation of $\langle E \rangle$. As a consequence, if the spectral assumption is changed, $F_{\rm max}$ remains constant and $\Phi_{\gamma, \, \rm S}$ is unchanged.  
Here, we use the best-fit values $L_{\rm max} =  5.1^{+3.4}_{-2.2} \times 10^{35}{\rm erg\;s^{-1}}$ and $\mathcal{N} = 18^{+14}_{-7}$ derived in \cite{Cataldo:2020qla} for $\alpha=1.5$ by fitting the flux, longitude and latitude distributions of the sample of 32 HGPS sources above the H.E.S.S. completeness threshold. 

\section{Total neutrino source flux}
\label{App:neutrino source}

The neutrino source flux is obtained from the $\gamma$-ray flux following the approach of \cite{Vecchiotti:2023nqy}.
The CR-injected spectrum is parameterized as a power law with an exponential cutoff $\phi_{\rm p}(E) \propto E^{-\Gamma_{\rm p}} \, \exp\left(-E/E_{\rm cut}\right)$. 
The proton spectral index is fixed to $\Gamma_{p} = 2.4$ to reproduce the average spectral properties of HGPS sources. 
The proton cutoff energy varies in the range $E_{\rm cut}=0.5-10$~PeV to explore the relevance of this parameter for our final results.   
%
The all-flavor neutrino spectrum (normalized in the 1-100 TeV energy window) produced by hadronic interaction within the source is given by:
\begin{eqnarray} \label{gamma and neutrino source flux}
\phi_{\nu}(E_{\nu};\,E_{\rm cut}) &=&  \frac{1}{K_\nu(E_{\rm cut})}\sum_{l=e,\mu,\tau}
\int_{E_{\nu}}^{\infty} dE\, \frac{d
\sigma_{l}(E,E_{\nu})}{dE_{\nu}}
\phi_{p}(E;\,E_{\rm cut})
\end{eqnarray}
where $K_\nu(E_{\rm cut})$ is the normalization constant: 
\begin{eqnarray} \label{gamma and neutrino nromalizations}
K_\nu (E_{\rm cut}) &=&  \sum_{l=e,\mu,\tau} \int^{E_{\rm sup}}_{ E_{\rm inf}} dE_\nu
\int_{E_{\nu}}^{\infty}  dE\, \frac{d
\sigma_{l}(E,E_{\nu})}{dE_{\nu}} \, \phi_{p}(E ; \, E_{\rm cut})
\end{eqnarray}
where $E_{\rm inf} = 1\,{\rm TeV}$ and $E_{\rm sup} = 100\,{\rm TeV}$. 
By using Eq.~\ref{gamma and neutrino source flux} and Eq.~\ref{phitot}, we calculate the cumulative neutrino emission produced by all sources (resolved and unresolved) contained in a given OW.
The all-flavor differential neutrino flux is given by:
\begin{equation}
\varphi_{\nu,\, s} \left(E_\nu; \, E_{\rm cut}, \, \xi \right) = \xi \, \Phi^{\rm max}_{\nu, \rm s}(E_{\rm cut}) \,  \phi_\nu (E_\nu;E_{\rm cut})
\end{equation}
where $\Phi^{\rm max}_{\nu, {\rm s}} \equiv \eta \, \Phi_{\gamma,\rm s}$. 
The parameter $\eta$ represents the ratio between the number of neutrinos (of all flavor) and the number of photons that a given source produce in the energy window 1-100 TeV and it is defined as $\eta \equiv \frac{K_{\nu}}{K_{\gamma}}$ where:
\begin{equation}
K_\gamma (E_{\rm cut}) = \int^{E_ {\rm sup}}_{E_{\rm inf}} dE_\gamma
\int_{E_{\gamma}}^{\infty}  dE\,
\frac{d\sigma(E,E_{\gamma})}{dE_{\gamma}}\,
\phi_{p}(E; \, E_{\rm cut})
\end{equation}
The flux $\Phi^{\rm max}_{\nu, {\rm s}}$ represents the maximal neutrino source contribution, i.e. the one obtained by assuming that all the TeV $\gamma$-ray sources, resolved and unresolved, are powered by hadronic processes.
We introduce the quantity $\xi\le 1$ to consider the possibility that only a fraction of the $\gamma$-ray source flux is produced by hadronic interaction and, hence, is accompanied by neutrino production.

\bibliographystyle{aasjournal}
\bibliography{bibliography.bib}

\end{document}